\documentclass[reprint, twocolumn, showkeys]{revtex4}

\usepackage{graphicx}% Include figure files
\usepackage{dcolumn}% Align table columns on decimal point
\usepackage{bm}% bold math
\usepackage{color}% bold math
\usepackage{amsmath,amsthm}
\usepackage{amssymb,bm}
\usepackage{mathrsfs}
\usepackage{comment}
\usepackage{ulem}

\graphicspath{%
    {converted_graphics/}% inserted by PCTeX
    {/}% inserted by PCTeX
}
\begin{document}

\title{Superreaction: the collective enhancement of a reaction rate by molecular polaritons in the presence of energy fluctuations}
\author{Nguyen Thanh Phuc}
\email{nthanhphuc@moleng.kyoto-u.ac.jp}
\affiliation{Department of Molecular Engineering, Graduate School of Engineering, Kyoto University, Kyoto 615-8510, Japan}

%%%%%%%%%%%%%%%%%%%%%%%%%%
\begin{abstract}
Recent experiments have demonstrated that molecular polaritons, hybrid states of light and matter formed by the strong coupling between molecular electronic or vibrational excitations and an optical cavity, can substantially modify the physical and chemical properties of molecular systems. Here, we show that, by exploiting the collective character of molecular polaritons in conjunction with the effect of polaron decoupling, i.e., the suppression of environmental influence on the polariton, a superreaction can be realized, involving a collective enhancement of charge or excitation-energy transfer reaction rate in a system of donors all coupled to a common acceptor. This effect is analogous to the phenomenon of superradiation. Since the polariton is a superposition state of excitations of all the molecules coupled to the cavity, it is vulnerable to the effect of decoherence caused by energy fluctuations in molecular systems. Consequently, in the absence of a strong light-matter interaction, the reaction rate decreases significantly as the number of molecules increases, even if the system starts from the polariton state. By turning on the light-matter interaction, the dynamic behavior of the system changes dramatically, and the reaction rate increases with the number of molecules, as expected for a superreaction. The underlying mechanism is shown to be the enhancement of quantum coherence between different donors as the light-matter interaction becomes stronger.
\end{abstract}


\maketitle

%%%%%%%%%%%%%%%%%%%%%
\textit{Introduction--}
The last decade has witnessed the emergence of a new field of study around molecular polaritons.
Polaritons modify the physical and chemical properties of molecular systems significantly through the strong coupling of electronic or vibrational molecular excitations to an optical cavity. 
This coupling leads to the formation of a hybrid state of light and matter~\cite{Ebbesen16, Ribeiro18, Feist18, Hertzog19, Herrera20}, resulting in various interesting phenomena.
Important applications have been proposed and demonstrated, including the manipulation of chemical landscapes~\cite{Hutchison12, Herrera16, Galego16, Takahashi19}, the modification of chemical reactivity by molecular-vibration polaritons~\cite{Thomas16, Thomas19, Hiura18, Lathera19, Galego19, Angulo19, Phuc20, Li20}, cavity-enhanced energy transfer and conductivity in organic media~\cite{Feist15, Schachenmayer15, Orgiu15}. 
Further applications include polariton lasing and Bose-Einstein condensates~\cite{Cohen10, Cwik14, Plumhof14, Lerario17}, and nonlinear optical properties with applications in optoelectronic devices~\cite{Barachati18, Jayaprakash19}.

In the absence of interaction with the environment, the molecular polariton has the form of a superposition state in which all the molecules coupled to the cavity are collectively excited. 
However, recent studies have shown that, in most cases, the collective effect displayed by a molecular polariton does not prominently influence chemical reactivity, except for its contribution to the collective coupling strength between the polariton and the cavity. 
In some cases, no collective effect is observed~\cite{Galego19, Angulo20}; the collective coupling can even lead to the suppression of an effect that scales as $1/N$, where $N$ is the number of molecules~\cite{Phuc20}. 
Therefore, it is unclear whether the inherent collective excitation in the polariton state can play a key role in chemical reactions or not. 
On the other hand, the collective coupling reportedly produces dramatic changes in physical system properties. 
A notable example is the phenomenon of superradiance~\cite{Dicke54, Gross82}, where the collective interaction of a group of emitters with a vacuum field causes them to emit a short burst of light radiation with a strong intensity proportional to $N^2$. 
This behavior differs drastically from the conventional exponential decay of a group of independent emitters with a rate proportional to $N$. 
Superradiance has been observed in systems ranging from quantum-dot arrays~\cite{Scheibner07} to molecular aggregates~\cite{Spano89, Deboer90, Fidder90, Meinardi03}.  

In this study, we exploit the inherent collective character of the molecular polariton state in a system of donors all coupled to a common acceptor, to obtain a superreaction with a collective enhancement of charge or excitation-energy transfer reaction rate.
This effect is analogous to the superradiance described above.
The collective enhancements that occur in superradiance and superreaction are purely quantum mechanical effects resulting from the constructive quantum interference between pathways starting from different group members. 
Thus, quantum coherence between donors is a necessary condition for realizing superreaction. 
On the other hand, the energy fluctuations caused by the molecule-environment interactions make the molecular polariton state vulnerable to decoherence. 
Therefore, a collective enhancement of the reaction rate is not guaranteed even if the system starts from the molecular polariton state. 
We numerical simulated molecular quantum dynamics, subject to system-environment interactions, to show that energy fluctuations decrease the reaction rate considerably as the number of molecules increases. 
Attaining a superreaction appears to be impossible in a system of molecules in condensed phase, where environmental effects cannot be ignored. 
However, a sufficiently strong light-matter interaction between molecules and the cavity causes a dramatic change in the system dynamics.
In this case, the reaction rate increases with the number of molecules coupled to the cavity, as expected for a superreaction. We show that the underlying mechanism of this change in behavior is the enhancement of quantum coherence between donors as the light-matter interaction becomes stronger. 
The effect of polaron decoupling is thus demonstrated in a dynamic context.
This is in contrast to its static manifestation in optical spectroscopy, which was first studied for a molecular aggregate with electronic excitation coupled to a single vibration mode~\cite{Spano15}. 
The effect of polaron decoupling was then generalized to the case where the electronic excitation was coupled to an environment consisting of many vibration modes spanning a wide frequency range ~\cite{Phuc19}.
It has been observed experimentally~\cite{Takahashi20}. 

\begin{figure}[tbp] % float placement: (h)ere, page (t)op, page (b)ottom, other (p)age
  \centering
  % file name: E:/Draft-Superreaction (Mar 2021)/Fig1.eps
  \includegraphics[keepaspectratio]{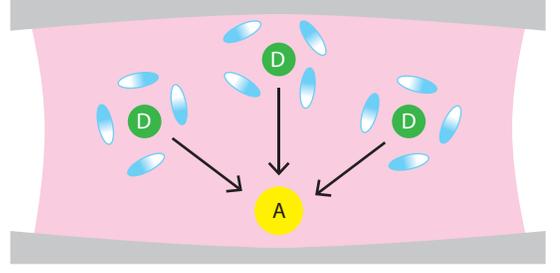}
  \caption{Schematic illustration of a superreaction, involving the collective enhancement of the electron or exciton transfer reaction rate in a system of donors (green) coupled to a common acceptor (yellow), all located inside an optical cavity. 
The collective coupling between the molecules and the cavity field (light magenta) produces a molecular polariton, a hybrid state of light and matter. 
The blue ellipsoids surrounding the donors represent the molecular environments whose thermal dynamics give rise to strong energy fluctuations in the molecular system.}
  \label{fig: system}
\end{figure}

As illustrated in Fig.~\ref{fig: system}, we consider an electron or excitation-energy (exciton) transfer reaction from a system of $N$ identical donors coupled to a common acceptor, modeled by the Hamiltonian $\hat{H}_\text{tr}=\hat{H}_\text{D}^\text{tr}+\hat{H}_\text{A}+\hat{H}_\text{DA}$ with
\begin{align}
\hat{H}_\text{D}^\text{tr}=\sum_{j=1}^N 
\hbar\omega_\text{D}|\text{e}_j\rangle\langle \text{e}_j|,
\end{align} 
\begin{align}
\hat{H}_\text{A}=\left\{\hbar\omega_\text{A}+\sum_{j=1}^N \sum_\xi g^\text{DA}_\xi 
\left[\left(\hat{b}^j_\xi\right)^\dagger+\hat{b}^j_\xi\right]\right\}
|\text{a}\rangle\langle \text{a}|
\end{align}
and
\begin{align}
\hat{H}_\text{DA}=-\sum_{j=1}^N \hbar V_\text{DA}
\left(|\text{a}\rangle\langle \text{e}_j|+|\text{e}_j\rangle\langle \text{a}|\right).
\label{eq: donor-acceptor coupling}
\end{align}
Here, $|\text{e}_j\rangle$ denotes the one-exciton state with the $j$th donor being in its electronic excited state, while the other molecules are in their electronic ground states; 
$|\text{a}\rangle$ is the state where the electron or exciton is transferred to the acceptor (in the case of electron transfer, the resulting positive charge is assumed to be distributed equally among donors); 
$\hbar\omega_\text{D}$ and $\hbar\omega_\text{A}$ are the energies of, respectively, a donor in its electronic excited state and the acceptor;
$V_{DA}$ is the coupling strength between the donors and the acceptor. 
The transfer from each donor to the acceptor is assumed to be accompanied by a configurational change in an independent environment, which is modeled by a collection of harmonic oscillators with the Hamiltonian $\hat{H}_\text{env}^\text{tr}=\sum_{j=1}^N \sum_\xi \hbar\omega_\xi \left(\hat{b}^j_\xi\right)^\dagger\hat{b}^j_\xi$. 
The reorganization energy associated with the transfer is given by $\lambda_\text{DA}=\sum_\xi \left(g_\xi^\text{DA}\right)^2/(\hbar\omega_\xi)$.

The donor electronic excitations are all coupled to a single cavity mode, as specified by the interaction Hamiltonian
\begin{align}
\hat{H}_\text{I}=-\sum_{j=1}^N \hbar\Omega_\text{R}
\left(|\text{e}_j\rangle\langle \text{g}|\hat{c}+|\text{g}\rangle\langle \text{e}_j|\hat{c}^\dagger\right),
\end{align}
where $|\text{g}\rangle=\prod_{j=1}^N \otimes |\text{g}_j\rangle$ is the ground state of the system with all the donors in their electronic ground states; 
$\hat{c}$ is the annihilation operator of a cavity photon; 
and $\Omega_\text{R}$ is the single-emitter Rabi frequency that characterizes the light-matter interaction strength. 
Here, a rotating-wave approximation is applied, as we do not consider an ultrastrong coupling regime~\cite{Phuc20}. 
The Hamiltonian of the cavity photons is given by $\hat{H}_\text{ph}=\hbar\omega_\text{c}\hat{c}^\dagger\hat{c}$, where $\omega_\text{c}$ is the cavity resonance frequency. 
Below, we consider resonant excitation, namely, $\omega_\text{c}=\omega_\text{D}$. 
The electronic excitation of each donor is assumed to be accompanied by a change in the configuration of an independent environment, as modeled by the coupling Hamiltonian
\begin{align}
\hat{H}_\text{D}^\text{ex}=\sum_{j=1}^N \sum_\chi g^\text{ex}_\chi 
\left[\left(\hat{b}^j_\chi\right)^\dagger+\hat{b}^j_\chi\right]
|\text{e}_j\rangle\langle \text{e}_j|
\end{align}
and the environment Hamiltonian $\hat{H}_\text{env}^\text{ex}=\sum_{j=1}^N \sum_\chi \hbar\omega_\chi \left(\hat{b}^j_\chi\right)^\dagger\hat{b}^j_\chi$. 
The reorganization energy associated with electronic excitation is given by $\lambda_\text{ex}=\sum_\chi \left(g_\chi^\text{ex}\right)^2/(\hbar\omega_\chi)$. 

Consider a system initially prepared in the molecular polariton state
\begin{align}
|\text{p}\rangle=\frac{1}{\sqrt{2}}
\left(
|1\rangle\otimes|\text{g}\rangle+|0\rangle\otimes \frac{1}{\sqrt{N}}\sum_{j=1}^N |\text{e}_j\rangle\right),
\label{eq: molecular polariton state}
\end{align}
where $|0\rangle$ and $|1\rangle$ denote the photon-number states with zero and one cavity photon, respectively. 
We first make the \textit{a priori} assumption that environmental energy fluctuations negligibly affect the polariton state, so that the polariton state is maintained throughout the transfer reaction. 
Then, by Fermi's golden rule, the transfer rate in the weak-coupling and high-temperature limit is given by
\begin{align}
k_{\text{P}\to\text{A}}^\text{tr}=&
\sqrt{\frac{\pi}{\hbar^2 k_\text{B}T \lambda_\text{PA}}}
\left|\langle \text{p}|\hat{H}_\text{DA}|\text{a}\rangle\right|^2 \nonumber\\
&\times \exp \left[ -\frac{(\Delta E_\text{PA}-\lambda_\text{PA})^2}{4\lambda_\text{PA}k_\text{B}T}\right],
\end{align}
where $\Delta E_\text{PA}$ denotes the energy difference between the polariton and acceptor states, and $\lambda_\text{PA}$ is the reorganization energy associated with the transfer from the polariton to the acceptor. 
If the displacements of the potential energy surfaces associated with the electronic excitations in the system of $N$ donors is small compared with those associated with the transfer from the donors to the acceptor, we have an approximate relation $\lambda_\text{PA}\simeq N \lambda_\text{DA}$. 
On the other hand, Eqs.~\eqref{eq: donor-acceptor coupling} and \eqref{eq: molecular polariton state} give
\begin{align}
\langle \text{p}|\hat{H}_\text{DA}|\text{a}\rangle=-\hbar V_\text{DA}\sqrt{\frac{N}{2}}.
\end{align}
Therefore, the maximum transfer rate, obtained for $\Delta E_\text{PA}=\lambda_\text{PA}$, scales as $N^{1/2}$. 
On the other hand, if the donor system is independent and one donor is initially excited, the maximum transfer rate scales as $N^{-1/2}$. 
This factor is explained by the fact that, when the transfer is accomplished, changes in the environment configurations become associated with all of the donors, such that the reorganization energy $\lambda_\text{DA}$ is multiplied by $N$.

The above analysis suggests that a system initially prepared in the molecular polariton state can produce a superreaction whose rate is greater than that of a system of independent donors and increases with the number of molecules.
This is true provided that the influence of energy fluctuations from the environment on the polariton state can be ignored. 
On the other hand, because the molecular polariton is a superposition state involving collective excitation from all molecules coupled to the cavity (as expressed in Eq.~\eqref{eq: molecular polariton state}), it is conventionally argued that the polariton state is vulnerable to decoherence, namely the loss of quantum coherence due to energy fluctuations. 
Therefore, it is highly nontrivial whether a superreaction can be achieved or not. 
To examine the prediction of the above analysis, we numerically simulated the dynamics of the system described by the above set of Hamiltonians. 
To limit the computational workload (which increases exponentially with the number of independent environments), the environments associated with the electronic excitation of each donor and with the transfer from the donor to the acceptor are described by a common set of harmonic oscillators $\xi=\chi$. 
We also assumed that the coupling strengths are related by a constant $\eta$, that is, $g_\xi^\text{DA}=\eta g_\xi^\text{ex}\equiv \eta g_\xi$. 
The environmental dynamics are characterized by their correlation functions
\begin{align}
C(t)=\int_0^\infty \text{d}\omega\; 
J(\omega)
\left[\coth\left(\frac{\beta\omega}{2}\right)\cos\omega t -i\sin\omega t\right],
\end{align}
where $\beta=1/(k_\text{B}T)$ and $J(\omega)=\sum_\xi g_\xi^2 \delta(\omega-\omega_\xi)$ is the environmental spectral density. 
Considering the Drude-Lorentz spectral density $J(\omega)=2\lambda\tau\omega/(\tau^2\omega^2+1)$ in the high-temperature limit ($k_\text{B}T\tau/\hbar\gg 1$), the correlation function $C(t)$ becomes exponential. 
Here, $\lambda=\sum_\xi g_\xi^2/(\hbar\omega_\xi)$ is the reorganization energy, and $\tau$ is the relaxation time of the environment.
In this case, the dynamics of the reduced system can be obtained by integrating out the environmental degrees of freedom. 
The resulting hierarchical equation of motion describes the dynamics of open quantum systems for a wide range of coupling strengths~\cite{Tanimura06}. 
Our numerical simulation set the following parameter values for the molecular system and environment: $V_\text{DA}=10\;\text{cm}^{-1}$, $T=300\;\text{K}$, $\tau=250\;\text{fs}$, $\lambda=10\;\text{cm}^{-1}$, and $\eta=1$, which correspond to the high-temperature and strong-energy-fluctuation limits ($V_\text{DA}<\sqrt{\lambda k_\text{B}T}$). 
For the initial state, we consider the case in which the system is initially prepared in the molecular polariton state [Eq.~\eqref{eq: molecular polariton state}] by a vertical excitation under the Condon approximation, that is, the environments are at the equilibrium positions of the potential energy surfaces associated with the donor electronic ground states.

\begin{figure*}[tbp] % float placement: (h)ere, page (t)op, page (b)ottom, other (p)age
  \centering
  % file name: E:/Draft-Superreaction (Mar 2021)/Fig2.eps
  \includegraphics[keepaspectratio]{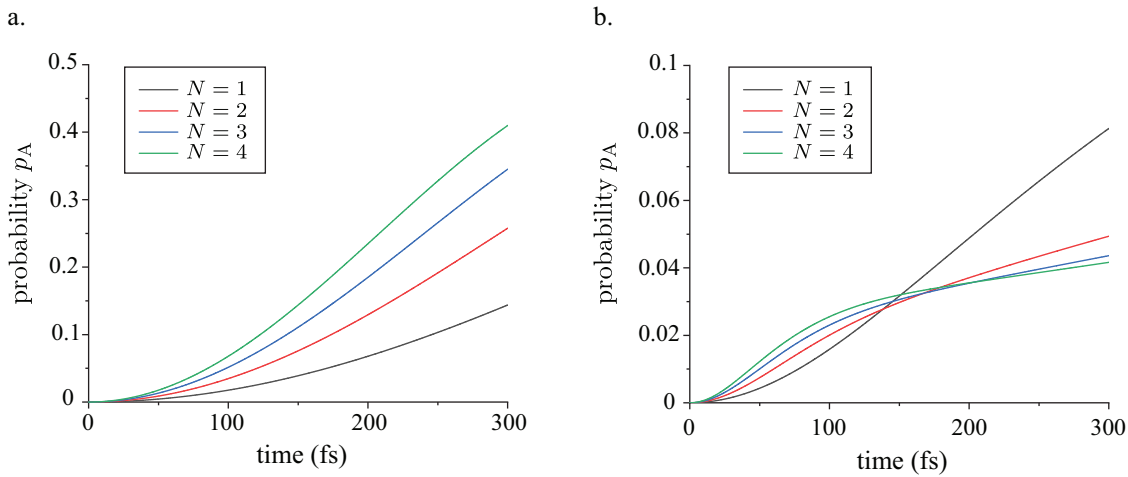}
  \caption{Comparison of electron/exciton transfer dynamics from a system of donors to an acceptor in the (a) absence and (b) presence of energy fluctuations when there is no coupling between the molecules and the cavity, namely $\Omega_\text{R}=0$.
The probability $p_\text{A}$ of finding an electron/exciton at the acceptor is plotted as a function of time for systems with different numbers $N$ of donors. 
In both cases, the system is initially prepared in the molecular polariton state given by Eq.~\eqref{eq: molecular polariton state}.
The parameters of the molecular system and environment are given in the text.}
  \label{fig: compare transfer dynamics between with and without energy fluctuation}
\end{figure*}

Figure~\ref{fig: compare transfer dynamics between with and without energy fluctuation} compares the transfer dynamics in the absence and presence of energy fluctuations if the molecules and the cavity are uncoupled, i.e., $\Omega_\text{R}=0$.
Here, the time-dependent probability $p_\text{A}(t)$ of finding an electron/exciton at the acceptor is calculated for systems with different numbers of donors $1\leq N \leq 4$. 
Evidently, despite the initial increase in transfer rate with the number of molecules due to the collective effect of the polariton state, the transfer rate rapidly changes its behavior as a result of energy fluctuations, becoming smaller for systems with many molecules. 
This change in behavior can be attributed to decoherence.

\begin{figure}[tbp] % float placement: (h)ere, page (t)op, page (b)ottom, other (p)age
  \centering
  % file name: E:/Draft-Superreaction (Mar 2021)/Fig3.eps
  \includegraphics[keepaspectratio]{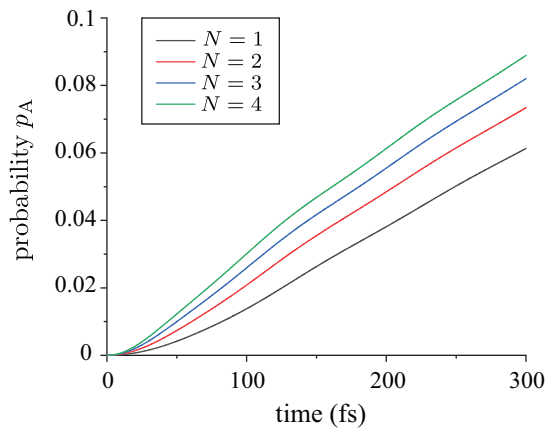}
  \caption{Transfer dynamics from a system of donors to an acceptor when the coupling between molecules and the cavity is turned on. 
As in Fig.~\ref{fig: compare transfer dynamics between with and without energy fluctuation}, the probability $p_\text{A}$ of finding an electron/exciton at the acceptor is plotted as a function of time for systems with different numbers $N$ of donors. 
The system is initially prepared in the molecular polariton state, and the dynamics occur in the presence of strong energy fluctuations. 
Here, the collective Rabi frequency is fixed at $\hbar\Omega_\text{R}\sqrt{N}=20\;\text{meV}$ as the number of molecules is varied.}
  \label{fig: transfer dynamics with Rabi frequency 20 meV}
\end{figure}

Figure~\ref{fig: transfer dynamics with Rabi frequency 20 meV} shows the transfer dynamics for systems with different numbers $N$ of molecules if the coupling between the molecules and the cavity is set to the collective Rabi frequency $\hbar\Omega_\text{R}\sqrt{N}=20\;\text{meV}$. 
Here, the single-emitter Rabi frequency $\Omega_\text{R}$ is changed while keeping the collective Rabi frequency fixed to the above value as $N$ is varied. 
The collective Rabi coupling is strong in the sense that it is greater than the energy fluctuation amplitude characterized by $\sqrt{\lambda k_\mathrm{B}T}$. 
It is clear from the numerical result that the transfer rate of transfer increases with $N$, as expected for a superreaction. 
By comparing this result with that in Fig.~\ref{fig: compare transfer dynamics between with and without energy fluctuation}b when the light-matter interaction is turned off, it can be inferred that the effect of decoherence is suppressed when the molecular system is strongly coupled to the cavity. 
This in turn leads to an increase in the transfer rate for systems with large $N$. 
This demonstrates the effect of polaron decoupling in a dynamic context, as opposed to its static manifestation in optical spectroscopy~\cite{Spano15, Phuc19, Takahashi20}.

\begin{figure*}[tbp] % float placement: (h)ere, page (t)op, page (b)ottom, other (p)age
  \centering
  % file name: E:/Draft-Superreaction (Mar 2021)/Fig4.eps
  \includegraphics[keepaspectratio]{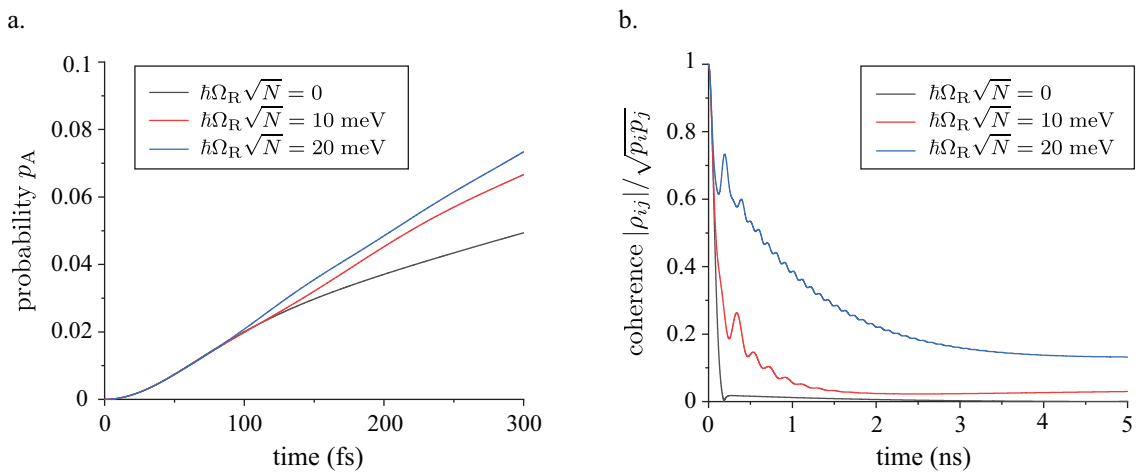}
  \caption{(a) Electron/exciton transfer dynamics from a system of donors to an acceptor, and (b) the quantum coherence between different donors for different values of the collective Rabi coupling strength $\Omega_\text{R}\sqrt{N}$. The quantum coherence between donors is quantified by $|\rho_{ij}(t)|/\sqrt{p_i(t)p_j(t)}$, where $\rho_{ij}\;(i\not=j)$ is the off-diagonal matrix element of the density matrix of the system, and $p_i, p_j$ are the probabilities of finding an electron/exciton at the $i, j$th donors.}
  \label{fig: Rabi-coupling-strength dependences of transfer dynamics and quantum coherence}
\end{figure*}

To justify the above prediction of the polaron decoupling mechanism for realizing a superreaction, we investigated the dependence on the Rabi coupling strength of the transfer dynamics and quantum coherence between different donors. 
Figure~\ref{fig: Rabi-coupling-strength dependences of transfer dynamics and quantum coherence} plots the time-dependent electron/exciton population $p_\text{A}(t)$ at the acceptor, and the quantum coherence between two different donors $i\not=j$, as quantified by the modulus of the off-diagonal matrix element $|\rho_{ij}(t)|/\sqrt{n_i(t)n_j(t)}$ of the density matrix of the system normalized by the populations, for different light-matter interaction strengths.
Evidently, the transfer rate increases with the coupling strength, and the underlying mechanism is the enhancement of the quantum coherence between different donors. 
This can be understood by noting that the collective enhancement is a purely quantum-mechanical effect, for which quantum coherence is an absolutely necessary condition.

In conclusion, we have shown that, by exploiting the collective character of the molecular polariton state, a superreaction with a collective enhancement of the electron/exciton transfer reaction rate can be obtained in a system of donors coupled to a common acceptor. 
Such a superreaction requires the persistence of quantum coherence between different donors in the presence of strong energy fluctuations.
This condition can be fulfilled by the effect of polaron decoupling, provided that the light-matter interaction is sufficiently strong. 
Finally, it is worth noting that the transfer dynamics considered in this study and their timescales depend on the coupling between the system and the environment as the system eventually approaches the thermal equilibrium state. 
Therefore, the dependence of the transfer rate with respect to the number $N$ of molecules coupled to the cavity, which is obtained by numerical simulations, should differ from the power scaling law obtained by the Fermi's golden rule analysis. 
Remaining avenues for future work include a quantitative investigation of the scaling power in the thermodynamic limit, that is, when $N\to\infty$, and in a nonequilibrium steady state.

%%%%%%%%%%%%%%%%%%%%%%%%%%
\begin{acknowledgements}
This work was supported by JSPS KAKENHI Grant Number 19K14638.
The computations were performed using Research Center for Computational Science, Okazaki, Japan.
\end{acknowledgements}

%%%%%%%%%%%%%%%%%%%%%%%%%%
\section*{Data availability}
The data that support the findings of this study are available from the corresponding author upon reasonable request.

%%%%%%%%%%%%%%%%%%%%%%%%%%

%%%%%%%%%%%%%%%

\begin{thebibliography}{100}
\bibitem{Ebbesen16}
T. W. Ebbesen,
"Hybrid Light-Matter States in a Molecular and Material Science Perspective,"
Acc. Chem. Res. \textbf{49}, 2403--2412 (2016).

\bibitem{Ribeiro18}
R. F. Ribeiro, L. A. Martinez-Martinez, M. Du, J. Campos-Gonzalez-Angulo, and J. Yuen-Zhou,
"Polariton chemistry: controlling molecular dynamics with optical cavities,"
Chem. Sci. \textbf{9}, 6325 (2018).

\bibitem{Feist18}
J. Feist, J. Galego, and F. J. Garcia-Vidal,
"Polaritonic Chemistry with Organic Molecules,"
ACS Photonics \textbf{5}, 205--216 (2018).

\bibitem{Hertzog19}
M. Hertzog, M. Wang, J. Mony, and K. Bojesson,
"Strong light-matter interactions: a new direction within chemistry,"
Chem. Soc. Rev. \textbf{48}, 937 (2019).

\bibitem{Herrera20}
F. Herrera and J. Owrutsky,
"Molecular polaritons for controlling chemistry with quantum optics,"
J. Chem. Phys. \textbf{152}, 100902 (2020).

\bibitem{Hutchison12}
J. A. Hutchison, T. Schwartz, C. Genet, E. Devaux, and T. W. Ebbesen, 
"Modifying Chemical Landscapes by Coupling to Vacuum Fields,"
Angew. Chem., Int. Ed. \textbf{51}, 1592--1596 (2012).

\bibitem{Herrera16}
F. Herrera and F. C. Spano,
"Cavity-Controlled Chemistry in Molecular Ensembles,"
Phys. Rev. Lett. \textbf{116}, 238301 (2016).

\bibitem{Galego16}
J. Galego, F. J. Garcia-Vidal, and J. Feist,
"Suppressing photochemical reactions with quantized light fields,"
Nat. Commun. \textbf{7}, 13841 (2016).

\bibitem{Takahashi19}
S. Takahashi, K. Watanabe, and Y. Matsumoto,
"Singlet fission of amorphous rubrene modulated by polariton formation,"
J. Chem. Phys. \textbf{151}(7), 074703 (2019).

\bibitem{Thomas16}
A. Thomas et al.,
"Ground-State Chemical Reactivity under Vibrational Coupling to the Vacuum Electromagnetic Field,"
Angew. Chem. Int. Ed. \textbf{55}, 11462--11466 (2016).

\bibitem{Thomas19}
A. Thomas et al.,
"Tilting a ground-state reactivity landscape by vibrational strong coupling,"
Science \textbf{363}, 615--619 (2019).

\bibitem{Hiura18}
H. Hiura, A. Shalabney, and J. George,
"Cavity Catalysis: Accelerating Reactions under Vibrational Strong Coupling,"
ChemRxiv, https://doi.org/10.26434/chemrxiv.7234721.v2 (2018).

\bibitem{Lathera19}
J. Lathera, P. Bhatta, A. Thomas, T. W. Ebbesen, and J. George,
"Cavity Catalysis by Cooperative Vibrational Strong Coupling of Reactant and Solvent Molecules,"
Angew. Chem. Int. Ed. \textbf{58}, 10635--10638 (2019).

\bibitem{Galego19}
J. Galego, C. Climent, F. J. Garcia-Vidal, and J. Feist,
"Cavity Casimir-Polder forces and their effects in ground state chemical reactivity,"
Phys. Rev. X \textbf{9}, 021057 (2019).

\bibitem{Angulo19}
J. A. Campos-Gonzalez-Angulo, R. F. Ribeiro, and J. Yuen-Zhou,
"Resonant catalysis of thermally-activated chemical reactions with vibrational polaritons,"
Nat. Comm. \textbf{10}, 4685 (2019).

\bibitem{Phuc20}
N. T. Phuc, P. Q. Trung, and A. Ishizaki,
"Controlling the nonadiabatic electron-transfer reaction rate through molecular-vibration polaritons in the ultrastrong coupling regime,"
Sci. Rep. \textbf{10}, 7318 (2020).

\bibitem{Li20}
T. E. Li, J. E. Subotnik, and A. Nitzan,
"Cavity molecular dynamics simulations of liquid water under vibrational ultrastrong coupling,"
Proc. Natl. Acad. Sci. USA \textbf{117}, 18324--18331 (2020).

\bibitem{Feist15}
J. Feist and F. J. Garcia-Vidal,
"Extraordinary Exciton Conductance Induced by Strong Coupling,"
Phys. Rev. Lett. \textbf{114}, 196402 (2015).

\bibitem{Schachenmayer15}
J. Schachenmayer, C. Genes, E. Tignone, and G. Pupillo,
"Cavity-Enhanced Transport of Excitons,"
Phys. Rev. Lett. \textbf{114}, 196403 (2015).

\bibitem{Orgiu15}
E. Orgiu et al.,
"Conductivity in organic semiconductors hybridized with the vacuum field,"
Nat. Mater. \textbf{14}, 1123--1129 (2015).

\bibitem{Cohen10}
S. Kena-Cohen and S. R. Forrest,
"Room-temperature polariton lasing in an organic single-crystal microcavity,"
Nat. Photonics \textbf{4}, 371--375 (2010).

\bibitem{Cwik14}
J. A. Cwik, S. Reja, P. B. Littlewood, and J. Keeling,
"Polariton condensation with saturable molecules dressed by vibrational modes,"
Euro. Phys. Lett. \textbf{105}, 47009 (2014).

\bibitem{Plumhof14}
J. D. Plumhof, T. Stoferle, L. Mai, U. Scherf, and R. F. Mahrt,
"Room-temperature Bose Einstein condensation of cavity exciton polaritons in a polymer,"
Nat. Mater. \textbf{13}, 247--252 (2014).

\bibitem{Lerario17}
G. Lerario et al.,
"Room-temperature superfluidity in a polariton condensate,"
Nat. Phys. \textbf{13}, 837 (2017).

\bibitem{Barachati18}
F. Barachati, J. Simon, Y. A. Getmanenko, S. Barlow, S. R. Marder, and S. Kena-Cohen,
"Tunable third-harmonic generation from polaritons in the ultrastrong coupling regime,"
ACS Photonics \textbf{5}(1), 119--125 (2018).

\bibitem{Jayaprakash19}
R. Jayaprakash, K. Georgiou, H. Coulthard, A. Askitopoulos, S. K. Rajendran, D. M. Coles, A. J. Musser, J. Clark, I. D. W. Samuel, G. A. Turnbull, P. G. Lagoudakis, and D. G. Lidzey,
"A hybrid organic-inorganic polariton LED,"
Light: Sci. Appl. \textbf{8}(1), 81 (2019).

\bibitem{Angulo20}
J. A. Campos-Gonzalez-Angulo and J. Yuen-Zhou,
"Polaritonic normal modes in transition state theory,"
J. Chem. Phys. \textbf{152}, 161101 (2020).

\bibitem{Dicke54}
R. H. Dicke,
"Coherence in spontaneous radiation processes,"
Phys. Rev. \textbf{93}, 99 (1954).

\bibitem{Gross82}
M. Gross and S. Haroche,
"Superradiance: an essay on the theory of collective spontaneous emission,"
Phys. Rep. \textbf{93}, 301-396 (1982).

\bibitem{Scheibner07}
M. Scheibner et al.,
"Superradiance of quantum dots,"
Nat. Phys. \textbf{3}, 106 (2007).

\bibitem{Spano89}
F. C. Spano and S. Mukamel,
"Superradiance in molecular aggregates,"
J. Chem. Phys. \textbf{91}, 683 (1989).

\bibitem{Deboer90}
S. Deboer and D. A. Wiersma,
"Dephasing-Induced Damping of Superradiant Emission in J-Aggregates,"
Chem. Phys. Lett. \textbf{165}, 45--53 (1990).

\bibitem{Fidder90}
H. Fidder, J. Knoester, and D. A. Wiersma,
"Superradiant Emission and Optical Dephasing in J-Aggregates,"
Chem. Phys. Lett. \textbf{171}, 529--536 (1990).

\bibitem{Meinardi03}
F. Meinardi, M. Cerminara, A. Sassella, R. Bonifacio, and R. Tubino,
"Superradiance in Molecular H Aggregates,"
Phys. Rev. Lett. \textbf{91}, 247401 (2003).

\bibitem{Spano15}
F. C. Spano,
"Optical microcavities enhance the exciton coherence length and eliminate vibronic coupling in J-aggregates,"
J. Chem. Phys. \textbf{142}, 184707 (2015).

\bibitem{Phuc19}
N. T. Phuc and A. Ishizaki,
"Precise determination of excitation energies in condensed-phase molecular systems based on exciton-polariton measurements,"
Phys. Rev. Research \textbf{1}, 033019 (2019).

\bibitem{Takahashi20}
S. Takahashi and K. Watanabe,
"Decoupling from a Thermal Bath via Molecular Polariton Formation,"
J. Phys. Chem. Lett. \textbf{11}, 1349--1356 (2020).

\bibitem{Tanimura06}
Y. Tanimura, 
"Stochastic Liouville, Langevin, Fokker-Planck, and master equation approaches to quantum dissipative systems,"
J. Phys. Soc. Jpn. \textbf{75}, 082001 (2006).

\end{thebibliography}
\end{document}